\documentclass[intlimits,twoside,a4paper]{article}

\usepackage[utf8]{inputenc}
\usepackage[eqsecnum,cm]{cmpj3}
\usepackage{subcaption}
\usepackage[version=4]{mhchem}
\newcommand{\vvec}[1]{\mathbf{#1}}

\articletype{Rapid Communication}
\title[Drug uptake and release in PNIPAM microgels]%
{Thermocontrolled drug uptake and release by a microgel via mesoscale simulations}
\author[O. Kalyuzhnyi, J. Ilnytskyi]%
{
Ostap Kalyuzhnyi\orcid{0000-0002-5691-3058}\refaddr{label1},
Jaroslav Ilnytskyi\orcid{0000-0002-1868-5648}\refaddr{label1,label2}
}
\addresses{
\addr{label1} Yukhnovskii Institute for Condensed Matter Physics of the National Academy of Sciences of Ukraine, 1 Svientsitskii Str., UA-79011 Lviv, Ukraine
\addr{label2} Institute of Applied Mathematics and Fundamental Sciences, Lviv Polytechnic National University, 5 Mytropolyta Andreia Str., UA-79013 Lviv, Ukraine}


\begin{document}
	
	\maketitle

\begin{abstract}
	We employ dissipative particle dynamics to study the uptake and release of a hydrophobic 
	therapeutic agent by a thermoresponsive microgel. Disordered polymer network of a microgel is 
	formed in explicit solvent by permanently crosslinking reactive sites on linear precursor chains, 
	and is subsequently equilibrated without further bond formation. We find that drug hydrophobicity 
	governs the balance between uptake and release: weak hydrophobicity limits uptake, whereas strong 
	hydrophobicity favors drug retention after microgel collapse. Within the present coarse-grained 
	framework, an intermediate drug hydrophobicity provides the most favorable compromise 
	between drug loading and release
	\printkeywords
\end{abstract}

\date{\today}

\section{\label{I}Introduction}

One of the challenges of drug-delivery systems is to ensure: (i) stability and diffusion of
drug-loaded carriers during transport stage and, (ii) effective drug release
at the intended site \cite{Chacko2012,Mura2013}. It can be addressed via stimuli-responsive carriers,
reacting either to endogenous stimuli (pH, redox conditions, etc.), or externally
applied stimuli (temperature, magnetic field, or light) \cite{Mura2013}. Stimuli-responsive microgels, soft
and permeable polymeric structures of $0.1$ to $10\mu m$ produced by crosslinking
polymer precursors, represent one of such options. Tunability of their size and porosity 
ensure controllability of the uptake, transport, and release of small drug molecules \cite{Karg2019,Plamper2017}.

Thermoresponsive microgels undergo structural rearrangement upon changes of a local
temperature, enabling a precise control of a drug release site  \cite{Schmaljohann2006,Bordat2019}.
The poly($N$-isopropylacrylamide) (PNIPAM) containing precursors are often used for this pupose.
The lower critical solution temperature (LCST) of PNIPAM in water is about $32^{\circ}\mathrm{C}$
\cite{Schild1992,Pelton2000}. Below it, the hydrogen bonding between its carbonyl and N-H 
groups and water dominate, leading to its hydrophilicity. Above LCST, the hydrophobic
association between isopropyl groups, and hydrogen bonding between neighboring amide groups
render PNIPAM chains hydrophobic resulting in the collapse of a microgel \cite{Futscher2017}.

Proximity of LCST to the physiological temperature makes such systems suitable for 
the thermocontrolled in-body delivery of small hydrophobic drugs, such as ibuprofen, aspirin, paracetamol,
doxorubicin, curcumin, etc. \cite{Mohan2024,Liu2014}. Uptake phase is characterised by the dominance of the
hydrophobic polymer-drug attraction ensuring stability of a drug-loaded carrier in water. The opposite is true during
the drug release phase, namely prevailance of the polymer-polymer attraction over the polymer-drug one and, as the
result, partial release of a drug into bulk.

Mesoscale simulations via dissipative particle dynamics (DPD) provide means to address this competition directly.
The system is split into soft spherical beads, each representing particular group of atoms, and the pairwise interaction between them is parametrised accordingly \cite{Groot1997}. The microgel can be formed by random crosslinling, instead of imposing an idealised homogeneous lattice \textit{a priory} \cite{Gnan2017}. The thermoresponse of a microgel can be traced then over sufficiently long times.
This approach is used in this study to examine how interaction parameters of the model affect the effectiveness of the drug uptake/release cycle.

\section{\label{II}DPD model, formation of microgel, drug uptake}

The model system in DPD approach comprises polymer, drug and water beads, encoded hereafter via $p$, $d$, $w$, respectively.
The total force acting between $i$-th and $j$-th beads contains conservative (C), dissipative (D), and random
(R), contributions\vspace{-1em}

\begin{equation}
	\label{eq:dpd_forces}
	\vvec F_{ij}
	=
	\vvec F^{\mathrm C}_{ij}
	+\vvec F^{\mathrm D}_{ij}
	+\vvec F^{\mathrm R}_{ij},
	\qquad
	\vvec F^{\mathrm C}_{ij}
	=
	\begin{cases}
		a_{ij}\left(1-r_{ij}\right)
		\hat{\vvec r}_{ij},
		& r_{ij}<1,\\
		0,
		& r_{ij}\geq 1,
	\end{cases}
\end{equation}
\vspace{-0.5em}

\noindent where $a_{ij}$ defines the repulsion strength between two beads, $r_{ij}$ is the separation between their centers of mass, and $\hat{\vvec r}_{ij}$ is the unit directed between their centers. The conservative force is of our main interest, whereas 
the other forces are treated as a thermostat, see details elsewhere \cite{Groot1997}. All the beads are of the diameter $\sigma=1$, which provides the length scale, and of mass, $m=1$. The energy unit is $k_{\mathrm B}T=1$. Total bead number density is set at $\rho=3$ \cite{Groot1997}.

The prototype for our model linear copolymer precursor is: GMA$_x$-NIPAM$_y$, where
GMA = \ce{C7H10O3} is the glycidyl methacrylat group, and NIPAM = \ce{C6H11NO} is the N-isopropylacrylamide one. For an
isotactic chain, the separation between the centers of adjacent groups can be estimated at about $2.6\text{\AA}$. On the other hand, persistent length for PNIPAM chain is about $5-8\text{\AA}$, therefore a single monomer of a flexible coarse-grained PNIPAM chain,
denoted as P, should contain at least 3 NIPAM groups. The monomer representing a single GMA group is denoted as G. We use the following combination of them as a precursor: G[P$_9$G]$_{5}$ resullting in 51-monomer chain. Given a typical bond length in
the DPD model is $0.9\sigma$,
the bead diameter is about $\sigma\approx 9\text{\AA}$ in real units. This value is of the same order that average dimensions of
a range of small drug molecules, such as ibuprofen, aspirin, paracetamol, doxorubicin, or curcumin, therefore these molecules
can be modelled as a single DPD bead. 

\begin{table}[htb]
	\centering
	\renewcommand{\arraystretch}{1.3}
		\begin{tabular}{|c|c|c|c|c|c|}
		\hline
		$T$ & $a_{ww}$=$a_{dd}$&$a_{pp}$&$a_{pw}$ & $a_{pd}$ & $a_{wd}$\\ \hline
		$T$<LCST & 25 & 25 & 25.6 & 25.6 & 30-80\\ \hline
		$T$>LCST & 25 & 22 & 38    & 27    & 30-80\\ \hline
	\end{tabular}
		\caption{DPD repulsion strength for the swollen state of
		the PNIPAM microgel at 	$T$<LCST, and its collapsed state at
		$T$>LCST. $p$, $w$, and $d$ denote polymer, water, and drug beads, respectively.}
		\label{tab:interactions}
\end{table}
Two characteristic cases are considered: $T<\mathrm{LCST}$ and $T>\mathrm{LCST}$. The repulsion strengths, $a_{ij}$, for all pairs of interacting beads in both cases, are collected in table~\ref{tab:interactions}. The water-water repulsion strength, $a_{ww}=25$, is obtained from fitting the compressibility of water at normal conditions \cite{Groot1997}. This is assumed for both below and above LCST, the same for the drug-drug, $a_{dd}$, interaction and the polymer-polymer, $a_{pp}$, one below LCST. Reduced value of $a_{pp}=22$ is used above LCST to promote collapse of a polymer in this case. The polymer-water repulsion strengths, $a_{pw}=25.6$ and $38$, respectively, are based on the temperature dependence of the Flory--Huggins mixing parameter, $\chi$ \cite{SotoFigueroa2012}, see Ref.~\cite{Yaremchuk2023}. The polymer-drug repulsion strength is set at $a_{pd}=a_{pw}=25.6$ below LCST,
 and $27$ above LCST to reflect some weakening of the polymer--drug affinity as found in the atomistic simulations of a PNIPAM--ibuprofen hydrogel \cite{Zeng2026}. Various possible levels of drug hydrophobicity, governed by $a_{wd}$, reflect properties
 of real small small hydrophobic drugs, these translates into the interval of $\chi=1.43-15.73$. 

The simulations are performed in a cubic box with $L_x=L_y=L_z=41$ with the periodic boundary conditions. Such size 
eliminates self-interaction of a swollen microgel with its periodic images. The initial system contains
$N_{\mathrm{mol}}=20$ linear precursors, with the structure as described above, in explicit solvent representing water.
The chains are initially placed in the central part of a cubic periodic simulation box to promote formation of a single
connected network. 

\begin{figure}[htb]
	\centering
	\includegraphics[angle=-90,width=0.6\linewidth]{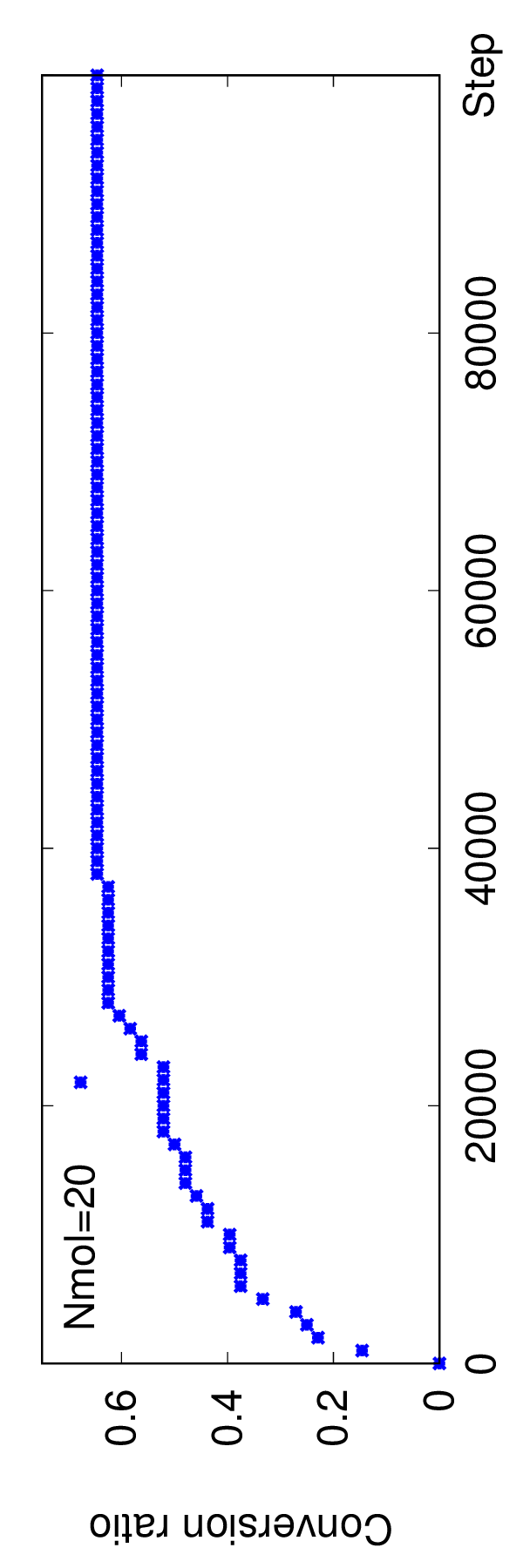}
\vspace{-1em}\\
	\caption{\label{CrosRatio} Evolution of the crosslinking conversion ratio during microgel
		formation. The network used in the uptake simulations is selected after the
		conversion has reached a plateau.}
\end{figure}

During the crosslinking stage, a permanent bond is created between two reactive beads when
they approach a separation of $r_{c}^{\mathrm{react}}=1.0$. The reactive beads are
monovalent. Crosslinking is performed for
$10^{5}$ DPD steps. The conversion curve reaches a plateau over this
interval, see figure~\ref{CrosRatio}, indicating that formation of more bonds became a very rare event if at all.
The largest connected polymer
cluster is identified as a microgel whereas disconnected subnets are excluded from the analysis. Microgel comprises
16 chains (816 polymer beads). It is equilibrated in explicit solvent with crosslink
formation disabled before drug beads are introduced.

\section{\label{III} Results}
After equilibration of a microgel in water, 191 water beads are converted into drug beads, yielding initial drug-to-polymer
loading of about $0.25$, on par with the typical experimental ibuprofen-to-polymer mass ratio \cite{Mohan2024}.

The spherically averaged density profile is evaluated for microgel, with respect to its center of mass, at each time
instance $t$. Then, we define the radius, $R_{\mathrm{mg}}$, of the microgel effective sphere, as the distance
where this profile turnes to zero. A drug bead is considered to be uptaken if it is located within this sphere, and
the number of uptaken drug is denoted by $N_d^{\mathrm{upt}}(t)$.

The drug uptake, occurring below LCST, is diffusion driven: the drug beads are able to enter the 
swollen microgel and are retained there by the hydrophobic interactions. This stage in
our simulations lasts for $t_1=1.5\cdot10^{6}$ DPD steps, whereas the release stage occurs
above LCST during the following time steps, from $t_1$ to $t_2=2\cdot10^{6}$. The adsorption capacity
at time $t$, $q(t)$, and delivery capacity, $p$, are defined as
\begin{equation}\label{qt_p}
	q(t)=
	\frac{N_d^{\mathrm{upt}}(t)}
	{N_p^{\mathrm{tot}}},
	\qquad
	p=q(t_1) - q(t_2),
\end{equation}
where $N_p^{\mathrm{tot}}$ is the total number of polymer beads belonging to a microgel. In an ideal scenario,
all drug beads are released, $q(t_2)=0$, and $p$ has a maximum value of $q(t_1)$. If $p<0$ then
the worst scenario occurs when drug uptake continues at the release stage. Both scenarios are separated by the boundary
case of $p=0$ when no drug release takes place. The values of $q(t_1)$ and $q(t_2)$ are evaluated within some time intervals
preceding these respective time instances, to improve statistics.

\begin{figure}[htb]
	\begin{center}
	\includegraphics[height=3.5cm]{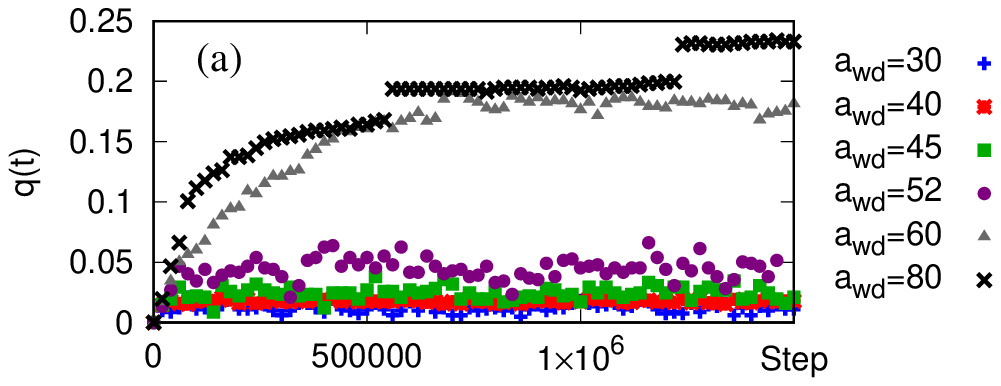}
	\includegraphics[height=3.5cm]{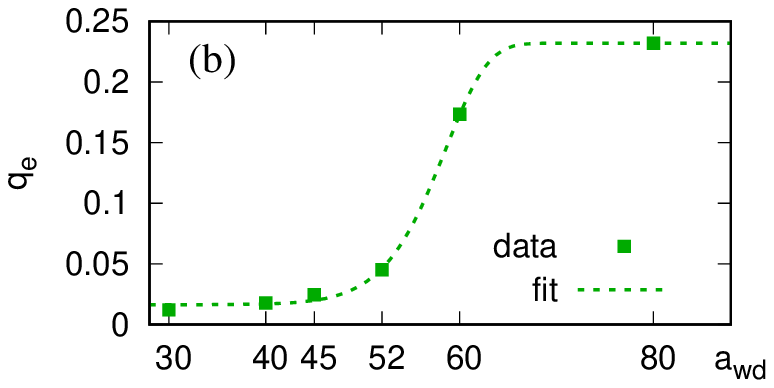}\vspace{-1.5em}\\
	\end{center}
    \vspace{-1em}
	\caption{\label{qt_evol_p}Drug uptake below LCST. (a) Time evolution of the adsorption capacity $q(t)$ (\ref{qt_p}). (b) Steady-state estimate $q_e \approx q(t_1)$ at various drug hydrophobicity, defined via $a_{wd}$. The data is fitted to the asymmetric logistic function $F_l(a_{wd})=d+(a-d)/[1+(a_{wd}/c)^b]^g$.}
	\label{fig:uptake_results}
\end{figure}

The time evolution of $q(t)$ during the uptake stage indicates the presence of a steady state with, $q_e \approx q(t_1)$, dependent
on the level of drug hydrophobicity given by $a_{wd}$, see figure~\ref{fig:uptake_results} (a).
For weakly hydrophobic drug ($a_{wd}$ is not far from $25$), $q_e$ is low. At elevated drug hydrophobicity (higher values of $a_{wd}$),
more drug is uptaken by a microgel manifested via higher valued of $q_e$. The dependence of $q_e$ on $a_{wd}$ is, however,
not linear but demonstrates a sharp increase at about $a_{wd}=52$, see figure~\ref{fig:uptake_results} (b). It 
can be fitted to the asymmetric logistic function $F_l(a_{wd})=d+(a-d)/[1+(a_{wd}/c)^b]^g$ with $a= 0.016$, $d=0.232$, $c=75.94$, $b=15.14$, and $g=46.59$, see figure~\ref{fig:uptake_results} (b). This steep increase of $q_e$, alongside with the step-wise jumps
observed for $q(t)$ evolution in figure~\ref{fig:uptake_results} (a), are discussed further below.

\begin{figure}[htb]
	\begin{center}
	\includegraphics[height=3.5cm]{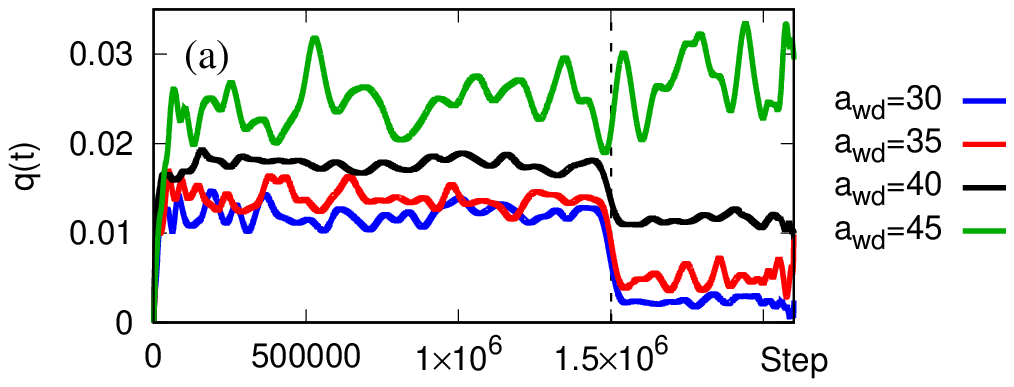}
	\includegraphics[height=3.5cm]{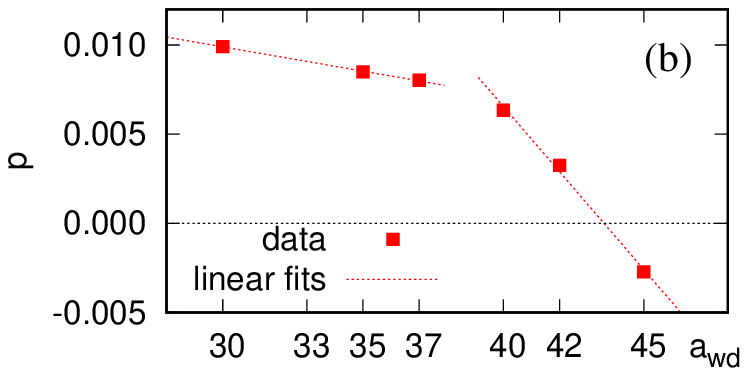}\vspace{-1.5em}\\
	\end{center}
    \vspace{-1em}
	\caption{(a) Time evolution of $q(t)$ during the drug uptake (up to $1.5\!\!\times\!\!10^6$) and
	 drug release (after $1.5\!\!\times\!\!10^6$) stages. (b) Delivery capacity $p$ (\ref{qt_p}) at various drug hydrophobicity, $a_{wd}$.
	 The results for $a_{wd}=40$ are averaged over 10 replicas.}
	\label{fig:release_drf}
\end{figure}

The drug release stage is initiated after $t_1=1.5\times10^{6}$ timesteps by switching all the relevant parameters
$a_{ij}$ to their respective values above LCST, according to table~\ref{tab:interactions}. The release stage is
conducted until the step $t_2=2.1\times10^{6}$, and it results in a microgel collapse into a compact desolvated object.
The drug release, however, does not occur at all levels of a drug hydrophobicity. Some representative cases are
shown in figure~\ref{fig:release_drf} (a), and a step-like decrease in $q(t)$ at $t>t_1$, indicating partial drug release
with the delivery efficiency $p$ (\ref{qt_p}), is observed at moderate drug hydrophobicity, $a_{wd}\leq 40$, only.
The dependence of $p$ on $a_{wd}$ is shown in figure~\ref{fig:release_drf} (b), and it indicates
some crossover-like behaviour here occuring at $a_{wd}\approx 39$, separating moderately and strongly
hydrophobic drugs, as marked via respective linear fits. At sufficiently strong drug hydrophobicity, $a_{wd}\geq45$,
$p$ turnes negative indicating uptaking more drug instead of releasing already captured. 
Thus, within the realms of a current model, only moderatively hydrophobic drugs with $a_{wd}=30-40$
can be released from the microgel interior.

\begin{figure}[htb]
	\centering
	\fbox{\includegraphics[width=0.22\linewidth]{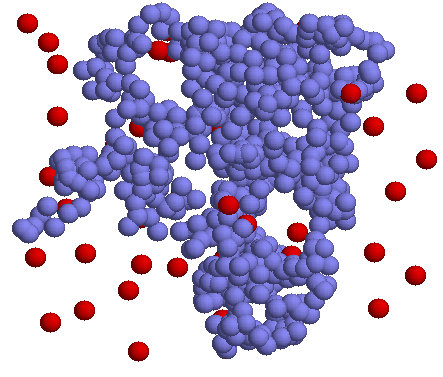}}
	\fbox{\includegraphics[width=0.22\linewidth]{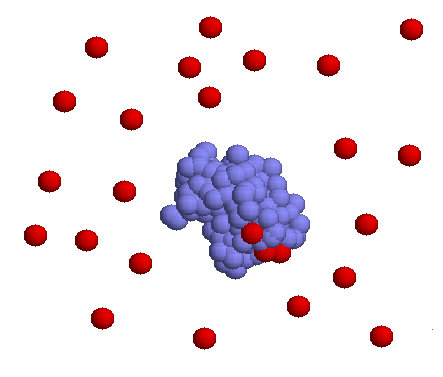}}\hspace{1em}
	\fbox{\includegraphics[width=0.22\linewidth]{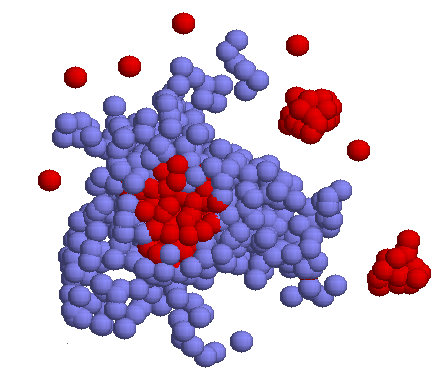}}
	\fbox{\includegraphics[width=0.22\linewidth]{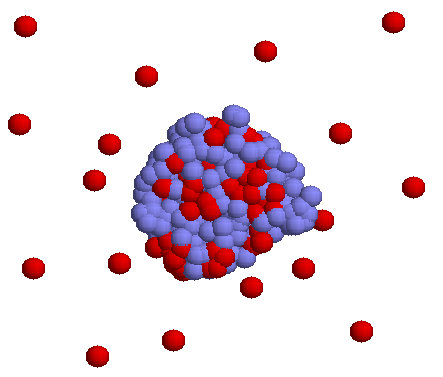}}\\
	\vspace{0.2em}
(a)\hspace{9em}(b)\hspace{10em}(c)\hspace{9em}(d)\\
	\caption{\label{snaps}Snapshots for a microgel (shown in blue) of 16 precursors and a drug (shown in red).\\
	(a) $a_{wd}=40$ uptake, (b) $a_{wd}=40$ release, (c) $a_{wd}=80$ uptake, (d) $a_{wd}=60$ release.}
\end{figure}

Such sharp difference between the release of moderately vs strongly hydrophobic drugs is explained by
their self-aggregation abilities in bulk. Typical snapshot of a moderately hydrophobic drug, $a_{wd}=40$, at the
end of an uptake stage is shown in figure~\ref{snaps} (a). We note that drug beads, shown in red, do not self-aggregate
neither in in bulk nor within a hydrogel in this case. At the end of a release stage, upon hydrogel collapse, part of
them are released into bulk one by one, see figure~\ref{snaps} (b). In contrary to this, a strongly hydrophobic drug,
$a_{wd}=60-80$, is seen to self-aggregate in bulk first, and then uptaken inside a gel in a form of such an aggregate,
see figure~\ref{snaps} (c). This explains the step-like evolution of $q(t)$ for $a_{wd}>60$, see figure~\ref{qt_evol_p} (a),
when each step is associated with an uptake of a single or more drug aggregates. Aggregates of strongly hydrophobic drugs
entrapped within a microgel, are more difficult to release, as shown in figure~\ref{snaps} (d).

Therefore, drug release above LCST is subject to the competition between two effects: (i) hydrophobic 
attraction between polymer beads drives microgel collapse and expulsion of both solvent and a drug from its interior,
favouring drug release, and (ii) hydrophobic drug self-aggregates and also is attracted to a hydrophobic polymer,
thus restricting drug release. For moderately hydrophobic drug, $a_{wd}=30-40$, the effect (i) dominates and delivery
capacity is positive, whereas for a strongly hydrophobic drug, $a_{wd}>40$, the effect (ii) is a dominant one and
it hampers drug release.

\section{\label{V}Conclusions}
We developed a coarse-grained model for a disordered PNIPAM-co-GMA like microgel, formed by explicit random
crosslinking of linear precursor chains in explicit solvent. The model reproduces microgel swollen and collapsed states through
temperature-dependent effective polymer--water and polymer--polymer interactions within the DPD method. Dimensional
analysis of coarse-graining leads to the estimate of the model bead diameter as $\sigma\approx 9\text{\AA}$.
As it is comparable with typical sizes of small drugs, such as: ibuprofen, aspirin, paracetamol, doxorubicin, curcumin, etc., these
are represented via a single bead. These drugs are characterized by various level of hydrophobicity, hence examination of
this effect on the uptake-release cycle, performed in this study, makes much sense.

Modelling of drug uptake is performed at the conditions mimicking the temperatures below LCST for the PNIPAM, when
a microgel is in a swollen state due to hydrogen bonding between its hydroxyl and amine groups and water, and drug
is attracted to hydrophobic groups of a microgel. These conditions are mimicked by appropriate choice of parameters
for effective interactions. With the increase of drug hydrophobicity, adsorption capacity demonstrates a non-linear behaviour,
linked to the self-aggregation of a drug. Step-like uptake dynamics is found in this case associated with uptake of drug
aggregates.

Modelling of drug release is performed at conditions reflecting the temperatures above the LCST, when the interchain hydrogen
bonding between amine and hydroxyl groups of PNIPAM dominate over their hydrogen bonding with water and, as the result,
a microgel collapses. Drug release is found to occur for the moderately hydrophobic drug only, when polymer-polymer hydrophobic attraction is stronger than the drug-polymer one. At certain level of drug hydrophobicity the crossover for the delivery capacity
(a mass of delivered drug per mass of a microgel) is observed to the regime of drug retention. Both increased drug-polymer attraction and
drug self-aggregation are responsible for strong retention of a drug in this case. The study reveals strong dependence of
the drug uptake-release cycle on the level of the drug hydrophobicity and the presence of some optimal interval
for the DPD repulsion parameter, $a_{wd}=30-40$, which translates into the Flory-Huggins mixing parameter of $\chi\approx 3.2-4.3$. Therefore, the microgel collapse alone does not guarantee drug release: a useful thermal cycle
requires a limited window of drug-water incompatibility. The work can be extended to cover other polymer precursors.

\section*{Acknowledgements}

This work was supported by the Grant of the President of Ukraine, provided by
the National Research Foundation of Ukraine, project No.~2025.05/0060,
``Prediction of the structure of thermoresponsive microgels by computer
simulation.''

\bibliographystyle{cmpj}
\bibliography{Micarogel_bib}
\ukrainianpart
\title{Мезоскопічне моделювання термоконтрольованого поглинання та вивільнення лікарських засобів}
\author{Остап Калюжний\refaddr{label1},
	Ярослав Ільницький\refaddr{label1,label2}}
\addresses{\addr{label1} Інститут фізики конденсованих систем імені Юхновського Національної академії наук України, вул. Свєнціцького, 1, UA-79011 Львів, Україна
	\addr{label2} Інститут прикладної математики та фундаментальних наук, Національний університет «Львівська політехніка», вул. Митрополита Андрея, 5, UA-79013 Львів, Україна
}

\makeukrtitle

\begin{abstract}
	
	Методом дисипативної динаміки досліджено поглинання та вивільнення гідрофобної лікарської речовини термочутливим мікрогелем. Невпорядковану полімерну сітку формували в явному розчиннику шляхом утворення зшивок між реакційноздатними центрами лінійних ланцюгів, після чого систему урівноважували без подальшого утворення зв’язків. Показано, що гідрофобність лікарської речовини визначає баланс між її поглинанням і вивільненням: низька гідрофобність обмежує поглинання, тоді як висока сприяє утриманню речовини після колапсу мікрогелю. Отже, у межах використаної огрубленої моделі проміжний рівень гідрофобності забезпечує найсприятливіший компроміс між завантаженням і вивільненням лікарської речовини.
	\keywords мікрогелі, доставка лікарських засобів, дисипатинва динаміка, ПНІПАМ, термочутливі полімери, гідрофобні лікарські речовини
\end{abstract}

\end{document}